\documentclass[epj]{webofc}
\usepackage[varg]{txfonts}   
\woctitle{QCD@Work 2016}
\usepackage{hyperref}
\usepackage[utf8]{inputenc}
\usepackage[T1]{fontenc}

\newcommand{\be}{\begin{equation}}
\newcommand{\ee}{\end{equation}}
\newcommand{\bea}{\begin{eqnarray}}
\newcommand{\eea}{\end{eqnarray}}
\newcommand{\bec}{\begin{center}}
\newcommand{\eec}{\end{center}}
\newcommand{\nn}{\nonumber}

\newcommand{\dd}{\displaystyle}

\begin{document}
\title{On the $h \to V \ell^+ \ell^-$ decays}

\author{\firstname{Pietro} \lastname{Santorelli }\inst{1,2}\fnsep\thanks{\email{pietro.santorelli@na.infn.it}} 
}

\institute{Dipartimento di Fisica "Ettore Pancini", Universit\`a  
di Napoli Federico II, Complesso Universitario di Monte Sant'Angelo,
Via Cintia, Edificio 6, I-80126 Napoli, Italy
\and
  INFN, Sezione di Napoli, I-80126 Napoli, Italy
  }

\abstract{%
A set of exclusive decay of the Higgs boson into a vector meson and a dilepton pair  ($h\to V \ell^+ \ell^-$, with $V=\Upsilon, J/\psi,\phi$, and $\ell=\mu, \tau$) are studied in the framework of the Standard Model. We have evaluated the decay rates, the dilepton mass spectra and the $V$  longitudinal helicity fraction distributions.  In the same framework, we considered the exclusive modes $h\to V \nu \bar \nu$ and  the implications of the CMS and ATLAS results for the lepton flavor-changing process  $h\to \tau^+ \mu^-$ on the $h\to V \tau^+ \mu^-$ decay modes.
}
\maketitle
\section{Introduction}
\label{intro}
The Higgs-like scalar observed at LHC with $m_h=125.7(4)$  GeV  
\cite{Aad:2012tfa,Chatrchyan:2012xdj,Agashe:2014kda} seems to fulfilled the prediction of the Standard Model (SM) 
for the Higgs boson. However, it is important to confirm that the couplings of the observed state to the fermions and 
gauge bosons are what the SM dictates. The couplings of the observed scalar to  top and beauty quarks  and to 
$\tau$ leptons are very well studied and are consistent with the SM predictions \cite{atlascms}; less known are the 
couplings to the light quarks and leptons. Many theoretical papers have been devoted to study how to modify these 
couplings as a consequences  of physic beyond the SM
 \cite{Contino:2013kra,Brivio:2013pma,Gonzalez-Alonso:2014eva,Gupta:2014rxa,Chien:2015xha}.

The study of the couplings to the first two generation of fermions is an experimental difficult job. From the theoretical point of view
the radiative  $h \to f \bar f \gamma$ processes have been considered with particular attention. 
 The leptonic modes $h \to \ell^+ \ell^- \gamma$ (with $\ell=e, \mu$) have been considered in 
 \cite{Abbasabadi:1996ze,Chen:2012ju,Sun:2013rqa,Dicus:2013ycd,Passarino:2013nka}. While the exclusive channels 
 $h \to V \gamma$, with $V$ a vector meson, have been scrutinized in \cite{Bodwin:2013gca, Kagan:2014ila,Isidori:2013cla,Koenig:2015pha}, 
and $h \to V Z$ have been studied in \cite{Bhattacharya:2014rra,Gao:2014xlv} as a way to measure the Higgs couplings to the light quarks.

Here we review the results obtained in \cite{Colangelo:2016jpi} where we have studied  the exclusive Higgs decays 
$h \to V \ell^+ \ell^-$, with $V=\Upsilon, J/\psi, \phi$ and $\ell$ is a light or a heavy charged lepton.
The motivations to study these processes are:
\begin{enumerate}
\item 
there is the possibility of considering, in addition to the decay rates, some distributions encoding important physical information, 
namely the distributions in the dilepton invariant mass squared;
\item 
due to the fact that several amplitudes contribute to each process, one can look at kinematical configurations 
where interferences are enhanced in order to get  information on the various Higgs couplings;
\item
they have a clear experimental signature, although the rates are small;
\item
deviations from the Standard Model can also be probed through the search of lepton flavor violating signals.  
\end{enumerate}

In the next section we will discuss the diagrams contributing to the $h \to V \ell^+ \ell^-$ processes, 
the couplings and the hadronic quantities necessary to evaluate them.
In section \ref{amplitudes} we show the calculation of the amplitudes, while, in section \ref{NumericalAnalysis},
the numerical results on the branching ratios, decay distributions and the fraction of polarized decay distribution
are presented and discussed. Finally, we will give our conclusions.
 
\section{The Relevant Diagrams}
\label{sec-1}
\begin{figure}
\centering
\begin{tabular}{ccc}
\includegraphics[width = .3\textwidth]{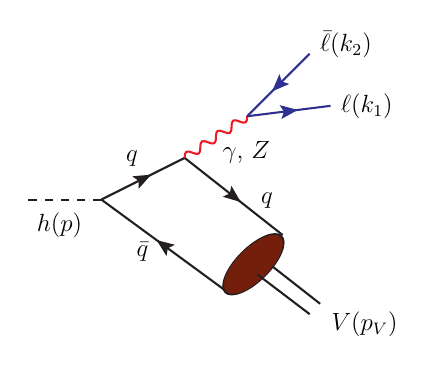} &
\includegraphics[width = .3\textwidth]{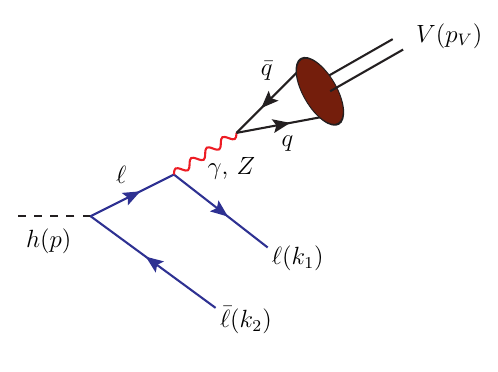} &
\includegraphics[width = .3\textwidth]{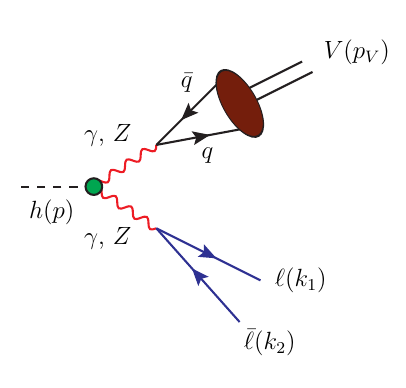}\\
(a)  &  (b) & (c) \\
\end{tabular}
\caption{\baselineskip 10pt Diagrams contributing to $h \to V \ell^+ \ell^-$ decays. In  (a) and (b) the diagrams with $\gamma$ and 
$Z$ emitted by $\bar q$ and $\bar \ell$ are considered. In (c)  the contributions of the $h \to ZZ$ and the effective 
$h \to \gamma \gamma$,  $h \to \gamma Z$ vertices are considered.}
\label{fig:diagrams}
\end{figure}

The amplitudes contributing to the $h \to V \ell^+  \ell^-$ decays  contain vertices in which the Higgs couples to quarks, to leptons and to 
the gauge bosons  $Z$ and $\gamma$.  In  SM such couplings are $g_{h f\bar f}=i\, m_f/v$ for 
fermions\footnote{For the quarks we use  the running masses evaluated at the Higgs mass scale $\mu\simeq m_h=125$ GeV at NNLO in the 
${\overline {\rm MS}}$ scheme.} and $g_{hZZ}=i\, 2m_Z^2/v$  for  $Z$ ($v=2 m_W/g=(\sqrt 2 G_F)^{-1/2} = 246$ GeV 
is the Higgs field vacuum expectation value).  Fig.~\ref{fig:diagrams} displays the three kinds of diagrams  that must be taken into account.

In the diagrams of Fig.~\ref{fig:diagrams} (a) we have the Higgs coupled to the quark-antiquark pair. 
The neutral gauge boson, $\gamma$ or $Z$,  is emitted from the quark or the antiquark line before they hadronize in the vector 
meson $V$. 
When the  dilepton invariant mass squared $q^2=(k_1+k_2)^2$ is low, the quark hadronization in the vector meson $V$ can be studied by 
using the formalism of the QCD hard exclusive processes \cite{Lepage:1979zb,Lepage:1980fj,Efremov:1979qk,Chernyak:1983ej}. 
Here we should consider the matrix element of the non-local quark antiquark operator while the vector meson state can be expressed as  an 
expansion in increasing twists, which  involves various  vector meson distribution amplitudes.
In the case of $V=\Upsilon,\,J/\psi,\,\phi$, the leading twist light-cone distribution amplitude (LCDA)  $\phi_\perp^V$  is defined from the matrix 
element of the non-local   ${\bar q}(y) \sigma_{\mu \nu} q(x)$ quark current:
 \be
\langle V(p_V ,\,\epsilon_V) | {\bar q}(y) \sigma_{\mu \nu} q(x)| 0 \rangle =-f_V^\perp (\epsilon_{V \, \mu}^* 
p_{V \, \nu} -\epsilon_{V \, \nu}^* p_{V \, \mu}) \int_0^ 1 du \, e^{i\,u\,p_V \cdot x+i\,{\bar u}\,p_V \cdot y} \phi_\perp^V(u)
\label{LCDA}
\ee
$({\bar u}=1-u)$.  $u\,p_V$  and ${\bar u}\,p_V$  represent the meson longitudinal momentum fraction  carried by the quark and 
antiquark.  $\phi_\perp^V$ is normalized to $1$; the hadronic parameter $f_V^\perp$ is discussed below. 
For  more details see \cite{Colangelo:2016jpi}.\\
The couplings of the Higgs to leptons are present in the diagrams of Fig.~\ref{fig:diagrams}~(b), with the $q \bar q$  pair emitted 
by the photon or  $Z$. Such diagrams  are  important  in the case of $\tau$. Here the hadronization of the $q \bar q$  
pair into the vector meson is represented by the matrix element
\be
\langle  V(p_V ,\,\epsilon_V) | {\bar q}\, \gamma_\mu\,  q| 0 \rangle=-i f_V m_V \epsilon^*_{V \,\mu} \,\, ,
\label{fV}
\ee
with $p_V$ and $\epsilon_V$ the $V$ meson momentum and polarization vector, respectively.
The experimental measurement of the decay width of $V \to e^+ e^-$ allow to measure the constant $f_V$. Less accessible 
is the hadronic parameter $f_V^\perp$  in \eqref{LCDA} and so results from lattice or QCD sum rule computations must 
be used.  In our analysis we use the  range for the ratio $R_{f_V}=f_V^\perp/f_V$ quoted in 
\cite{Grossmann:2015lea},  obtained exploiting  non-relativistic QCD  scaling relations \cite{Caswell:1985ui,Bodwin:1994jh}:

\be
f_\phi =0.223 \pm 0.0014 \,\,\, {\rm MeV}\,\,\, ,  \hskip 1.5cm R_{f_\phi}=0.76 \pm 0.04 \,\,, \nn \\
\ee
\be
f_{J/\psi} = 0.4033 \pm 0.0051 \,\,\, {\rm MeV}\,\,\, , \hskip 1cm R_{f_{J/\psi}}=0.91 \pm 0.14 \,\,, \label{constants} \\
\ee
\be
f_\Upsilon =0.6844 \pm 0.0046 \,\,\, {\rm MeV}\,\,\, , \hskip 1.3cm R_{f_\Upsilon}=1.09 \pm 0.04 \,\,. \nn
\ee

The diagrams of Fig.~\ref{fig:diagrams}~(c)  involve the coupling of the Higgs  to a pair of gauge bosons, which in turn are coupled to a  
lepton pair and to a $q \bar q$ pair  that hadronizes  into $V$.
The  elementary $hZZ$  coupling  can be obtained from the SM Lagrangian. The effective $h\gamma \gamma$ and $hZ\gamma$  vertices  
can be written as
\be
A(H \to  G_1 G_2)=i \frac{\alpha}{\pi v}C_{G_1 G_2}\left[g_{\mu \nu}(p_V \cdot q)-p_{V\mu} q_\nu \right]  \epsilon_{G_1}^{*\mu} \,\, \epsilon_{G_2}^{*\nu}  \,\,\, , \label{effective}
\ee
with $G_1$ and $G_2$  either $\gamma \gamma$ or $Z\gamma$,  and $\epsilon_{G_1}$, $\epsilon_{G_2}$   polarization vectors.
In Eq.~(\ref{effective}) $p_V$ is the momentum  of the meson V and $q$ the momentum of the dilepton.  The effective  
$h \gamma \gamma$ and $hZ\gamma$ couplings are determined by calculations of loop diagrams:  
$C_{\gamma \gamma}=-3.266 +i 0.021$ and $C_{\gamma Z}=-2.046 +i 0.005$ \cite{Koenig:2015pha}. 
In the $Z$ propagator we take into account, without considering the uncertainty,  the width $\Gamma(Z)=2.4952$ GeV \cite{Agashe:2014kda}. 
It is worth remarking that the possibility to access the $hZZ$ coupling is a feature of the class of modes we are analyzing. 
Moreover, since a sizeable contribution to $h \to V  \ell^+ \ell^-$  involves the effective  $h\gamma\gamma$ and $h Z \gamma$  couplings 
from diagrams  sensitive to  New Physics effects, the exclusive processes also probe deviations from SM.

\section{Decay Amplitudes}
\label{amplitudes}

In this section we give the expressions for the amplitudes corresponding to the diagrams of Fig.~\ref{fig:diagrams}.
At this end, we  define
\be
C_\gamma = 4 \pi \alpha Q_\ell Q_q \,\, , \hspace*{1cm}   
C_Z = \frac{4 \pi \alpha}{s_W^2 c_W^2}  \,,
\ee
with $s_W=\sin \theta_W$, $c_W=\cos \theta_W$, and $\theta_W$  the Weinberg angle, and write
the propagators in Fig.~\ref{fig:diagrams}  in terms of the  functions
\be
D_1(a,b, {\hat q}^2) = a+b {\hat q}^2-ab \, {\hat m}_V^2-{\hat m}_q^2\,, \hspace{0.9cm} 
D_2( {\hat q}^2) = {\hat q}^2- {\hat m}_Z^2+i \, {\hat m}_Z {\hat \Gamma}_Z\,, \hspace{0.9cm}
D_3({\hat k}) = 1-2 n \cdot {\hat k} \,, 
\ee
where $n=(1,{\vec 0})$ and we use  the notation ${\hat x}=\displaystyle{{x}/{m_h}}$, $x$ being a  mass or a momentum.
The lepton current, due to the intermediate gauge boson, has various Dirac structures,
\be
V_\ell^\mu  = {\bar \psi}_\ell (k_1) \gamma^\mu \psi_{\bar \ell}(k_2) \,,
\ee
for the photon; while for the intermediate $Z$ we have
\be
A_\ell^\mu  = {\bar \psi}_\ell (k_1) \gamma^\mu \gamma_5 \psi_{\bar \ell}(k_2)  \, ,
\hspace{1cm}
T_\ell^{\mu \nu} =  {\bar \psi}_\ell (k_1) \gamma^\mu \gamma^\nu \psi_{\bar \ell}(k_2) \, ,
\hspace{1cm}
{\tilde T}_\ell^{\mu \nu} =  {\bar \psi}_\ell (k_1) \gamma^\mu \gamma^\nu \gamma_5 \psi_{\bar \ell}(k_2) \, .
\ee
We write  the SM neutral current coupled to  the $Z$ boson as
\be
{\cal L}_\mu=\left( -\frac{ie}{s_W c_W} \right)\left(\Delta_V^f\, {\bar f}\gamma_\mu f + \Delta_A^f\, {\bar f}\gamma_\mu \gamma_5 f \right)\,,
\label{NC}
\ee
where $f$ denotes a fermion, and
\be
\Delta_V^f =\frac{1}{2} \left( T_3^f -2 s_W^2 Q^f \right) \, ,  \hspace*{1cm}
\Delta_A^f = -\frac{1}{2}  T_3^f  \, , \ee
with $ T_3^f$  the third component of the weak isospin and $Q^f$ the electric charge of   $f$.
Diagrams  in Fig.~\ref{fig:diagrams}(a) also involve the  integrals over the LCDA of the vector meson $V$:
\be
I_1 =I_1( {\hat q}^2)=\int_0^1\,du \, \phi_\perp^V(u)\left[\frac{1}{D_1(1-u,u,{\hat q}^2)}+\frac{1}{D_1(u, 1-u,{\hat q}^2)} \right]\,,\label{int1}\\
\ee
\be
I_2 = I_2( {\hat q}^2)=\int_0^1\,du \, \phi_\perp^V(u)\left[\frac{u}{D_1(1-u,u,{\hat q}^2)}+\frac{1-u}{D_1(u, 1-u,{\hat q}^2)} \right]\,.\label{int2}
\ee
We report the various expressions in correspondence with the diagrams in 
Fig.~\ref{fig:diagrams}(a), (b) and (c),  considering separately the intermediate photon and $Z$ contributions.
\begin{itemize}
\item Fig. \ref{fig:diagrams}(a), intermediate $\gamma$:
\be
A_{(a)}^\gamma=C_{(a)}^\gamma m_h \epsilon_V^{*\alpha} V_{\ell \, \mu} \, \left\{[n_\alpha {\hat p}_V^\mu- g_\alpha^\mu (n \cdot {\hat p}_V)] I_1-g^\mu_\alpha {\hat m}_V^2 I_2\right\}\,,
\hspace{0.6cm} {\rm with} \hspace{0.6cm}
C_{(a)}^\gamma=\frac{1}{m_h^2} \frac{\hat m_q}{v} C_\gamma  f_V^\perp  \frac{1}{\hat q^2}  \, .
\ee
\item  Fig.~\ref{fig:diagrams}(a),  intermediate $Z$:
\be
A_{(a)}^Z=C_{(a)}^Z \epsilon_V^{*\alpha}
\left[\Delta_V^\ell \, V_{\ell \, \mu}+ \Delta_A^\ell \, A_{\ell \, \mu}\right]\, \left( g^{\mu \alpha} p_V^\sigma -g^{\alpha \sigma} 
p_V^\mu \right)\,\left[ n_\sigma I_1-{\hat p}_{V \sigma} I_2 \right]\,,
\ee
with
\be
C_{(a)}^Z=-\frac{1}{m_h^2} \frac{\hat m_q}{v} C_Z \frac{1}{D_2({\hat q}^2)} f_V^\perp \,\Delta_V^q\,\,\, .
\ee
\item  Fig. \ref{fig:diagrams}(b),  intermediate $\gamma$:
\be
A_{(b)}^\gamma=C_{(b)}^\gamma \epsilon_V^{*\alpha} n^\mu \left[-\frac{1}{D_3({\hat k}_1)}\, T_{\ell \, \mu \alpha}+
\frac{1}{D_3({\hat k}_2)}\, T_{\ell \,  \alpha \mu} \right]\,,
\hspace{1cm} {\rm with} \hspace{1cm}
C_{(b)}^\gamma = \frac{1}{m_h^2} \frac{\hat m_\ell}{v} C_\gamma \frac{f_V m_V}{\hat m_V^2} \,.
\ee
\item Fig. \ref{fig:diagrams}(b), intermediate $Z$:
\be
A_{(b)}^Z=C_{(b)}^Z \epsilon^*_{V \alpha} n_\mu \left\{-\frac{1}{D_3({\hat k}_1)} \left[\Delta_V^\ell T_\ell^{\mu \alpha}+\Delta_A^\ell 
{\tilde T}_\ell^{\mu \alpha} \right] +\frac{1}{D_3({\hat k}_2)} 
\left[\Delta_V^\ell T_\ell^{ \alpha \mu}-\Delta_A^\ell {\tilde T}_\ell^{ \alpha \mu} \right] \right\}\,,
\ee
with
\be
C_{(b)}^Z=\frac{1}{m_h^2}\frac{\hat m_\ell}{v} C_Z \frac{\Delta_V^q}{D_2({\hat m}_V^2)}f_V m_V   \,\,\, .
\ee
\item  Fig. \ref{fig:diagrams}(c), two intermediate photons:
\be
A_{(c)}^{\gamma \gamma}=C_{(c)}^{\gamma \gamma} \epsilon_V^{*\alpha}[ g_{\alpha \mu} (q \cdot p_V)-m_h^2 n_\alpha n_\mu] \, V_{\ell}^{\mu}\,,
\hspace{1cm} {\rm with} \hspace{1cm}
C_{(c)}^{\gamma \gamma}=\frac{1}{m_h^4}\frac{\alpha}{\pi v} C_{\gamma \gamma} \, C_\gamma 
\frac{f_V m_V}{\hat m_V^2} \frac{1}{{\hat q}^2} \,\,\, .
\ee
\item Fig. \ref{fig:diagrams}(c),  two intermediate $Z$:
\be
A_{(c)}^{ZZ}=C_{(c)}^{ZZ}\epsilon^*_{V \alpha} \left(\Delta_V^\ell \,V_\ell^\alpha + \Delta_A^\ell \,A_\ell^\alpha \right)\,,
\hspace{0.6cm} {\rm with} \hspace{0.6cm}
C_{(c)}^{ZZ}=\frac{1}{m_h^2} \frac{2 {\hat m}_Z^2}{v}C_Z \frac{1}{D_2({\hat q}^2)}\frac{1} {D_2({\hat m}^2_V)} \Delta_V^q f_V m_V  \, .
\ee
\item  Fig. \ref{fig:diagrams}(c),  intermediate  $\gamma \, Z$, with  $\gamma$ converting to leptons:
\be
A_{(c)}^{\gamma Z}=C_{(c)}^{\gamma Z}\epsilon_V^{*\alpha}[ g_{\alpha \mu} (q \cdot p_V)-m_h^2 n_\alpha n_\mu] \, V_{\ell}^ {\mu}\,,
\hspace{0.5cm} {\rm with} \hspace{0.5cm}
C_{(c)}^{\gamma Z}=\frac{1}{m_h^4}\frac{\alpha}{\pi v}C_{\gamma Z} \frac{4 \pi \alpha Q_\ell}{s_W c_W} \frac{1}{{\hat q}^2} 
\frac{\Delta_V^q}{D_2({\hat m}^2_V)} f_V m_V \, .
\ee
\item  Fig. \ref{fig:diagrams}(c),  intermediate  $Z \, \gamma$, with $Z$ converting to leptons:
\be
A_{(c)}^{Z \gamma }=C_{(c)}^{Z \gamma }\epsilon_V^{*\alpha}[ g_{\alpha \mu} (q \cdot p_V)-m_h^2 n_\alpha n_\mu] \, 
\left(\Delta_V^\ell \,V_\ell^\mu + \Delta_A^\ell \,A_\ell^\mu \right)\,,
\ee
with
\be
C_{(c)}^{Z \gamma }=\frac{1}{m_h^4}\frac{\alpha}{\pi v}C_{\gamma Z} \frac{4 \pi \alpha Q_q}{s_W c_W}\frac{1}{{\hat m}^2_V}
 \frac{1}{D_2({\hat q}^2)} f_V m_V \,\,\, .
\ee
\end{itemize}
The effective couplings $C_{\gamma \gamma} $ and $C_{\gamma Z}$ are defined through Eq.~\eqref{effective}.

\section{Numerical Analysis}
\label{NumericalAnalysis}

Starting from the expressions obtained in the previous section we are able to compute the widths of the Higgs into the 
final states. But to calculate the branching fractions is necessary to get rid of the poorly known Higgs full width. 
One possibility is to use the expression  \cite{Koenig:2015pha}
\be
 {\cal B}(h \to V \ell^+ \ell^-) = \frac{\Gamma (h \to V \ell^+ \ell^-)}{\Gamma (h \to  \gamma \gamma)} {\cal B}(h \to \gamma \gamma)_{exp} \,\,\, 
 \ee 
which employs the computed  widths $\Gamma (h \to V \ell^+ \ell^-)$ and 
$\dd \Gamma (h \to \gamma \gamma)=\left(\alpha^2/(64 \pi^3 v^2)\right) |C_{\gamma \gamma}|^2 m_h^3$  combined with  
the measurement  
$\dd {\cal B}(h \to \gamma \gamma)_{exp}=(2.28 \pm 0.11 ) \times10^{-3}$ \cite{Heinemeyer:2013tqa}.  
We obtain
\bea
{\cal B}(h \to \phi \mu^+ \mu^-)& = & (7.93 \pm 0.39)  \times 10^{-8}  \hspace{0.6cm}
{\cal B}(h \to \phi \tau^+ \tau^-)=(2.35 \pm 0.12) \times 10^{-6}  \nn \\
{\cal B}(h \to J/\psi \mu^+ \mu^-)& = & (9.10 \pm 0.50) \times 10^{-8} \hspace{0.6cm} 
{\cal B}(h \to J/\psi \tau^+ \tau^-)=(1.82 \pm 0.10) \times 10^{-6}  \label{br} \\
{\cal B}(h \to \Upsilon \mu^+ \mu^-)& = &(5.60 \pm 0.37) \times 10^{-7} \hspace{0.6cm} 
{\cal B}(h \to \Upsilon \tau^+ \tau^-)=(5.66 \pm 0.29) \times 10^{-7}\,.  \nn
\eea

The errors in the branching ratios take into account the uncertainties on the LCDA parameters (cfr \cite{Colangelo:2016jpi}), on 
the decay constants $f_V$ and on the ratios $R_{f_V}$ in Eqs. (\ref{constants}), and the error on ${\cal B}(h \to \gamma \gamma)_{exp}$.
The largest contribution to the uncertainties on the branching ratios is due to the uncertainty on  ${\cal B}(h \to \gamma \gamma)_{exp}$ 
amounting to $50-60\%$ of the total error. The  uncertainty on $R_{f_V}$  constitutes $20-30\%$ of the total error. 

\begin{figure}
\centering
\includegraphics[width = .3\textwidth]{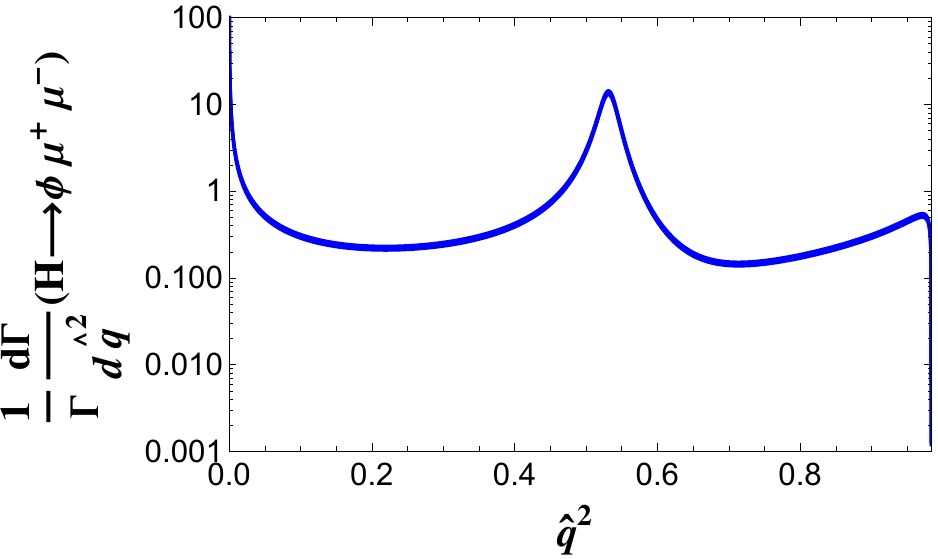}
\includegraphics[width = .3\textwidth]{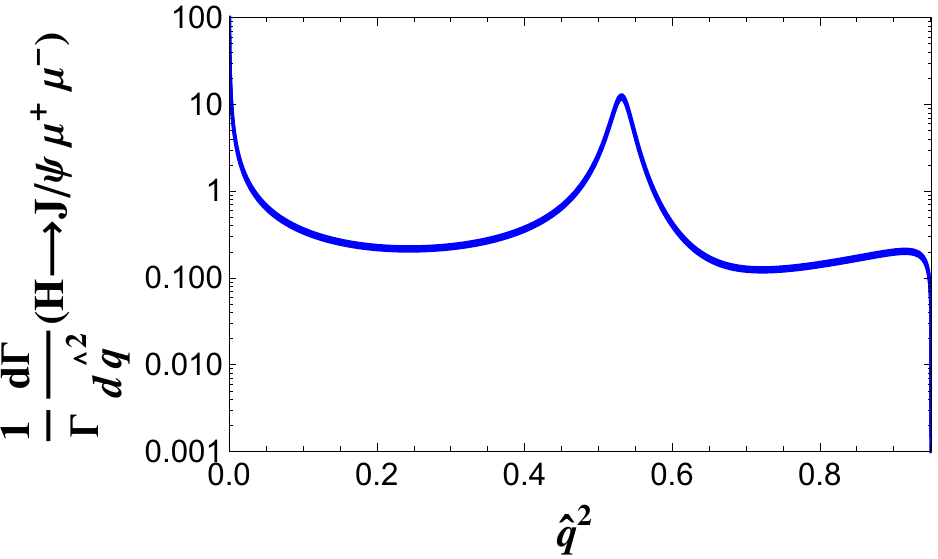}
\includegraphics[width = .3\textwidth]{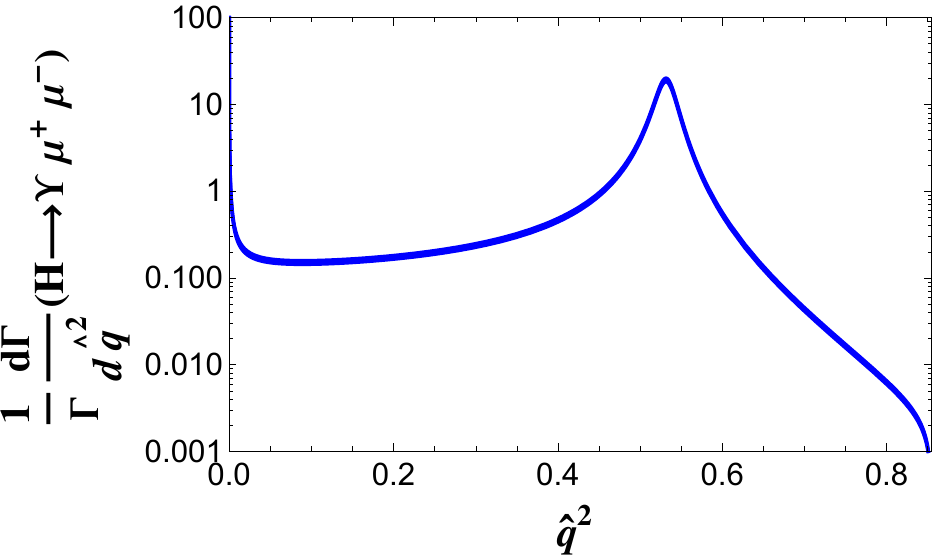}
\\
~\\
\includegraphics[width = .3\textwidth]{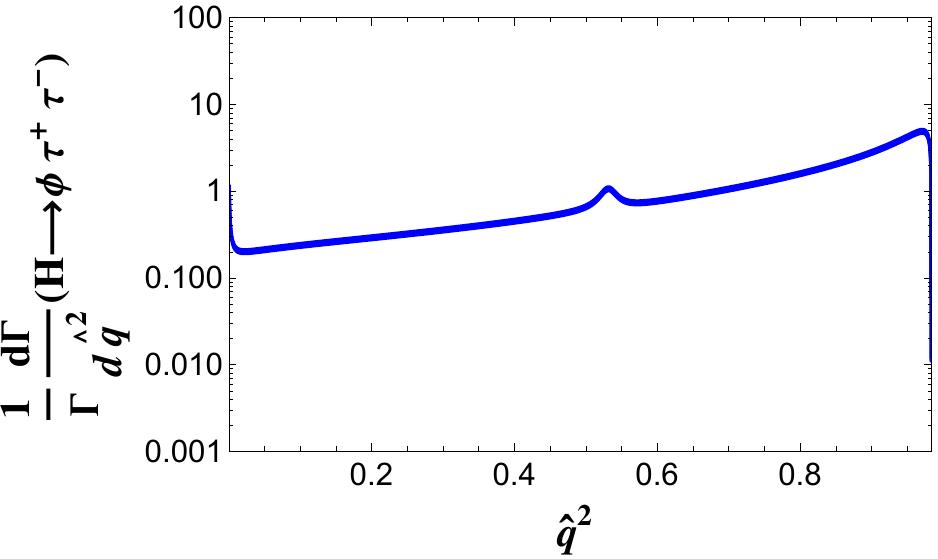}
\includegraphics[width = .3\textwidth]{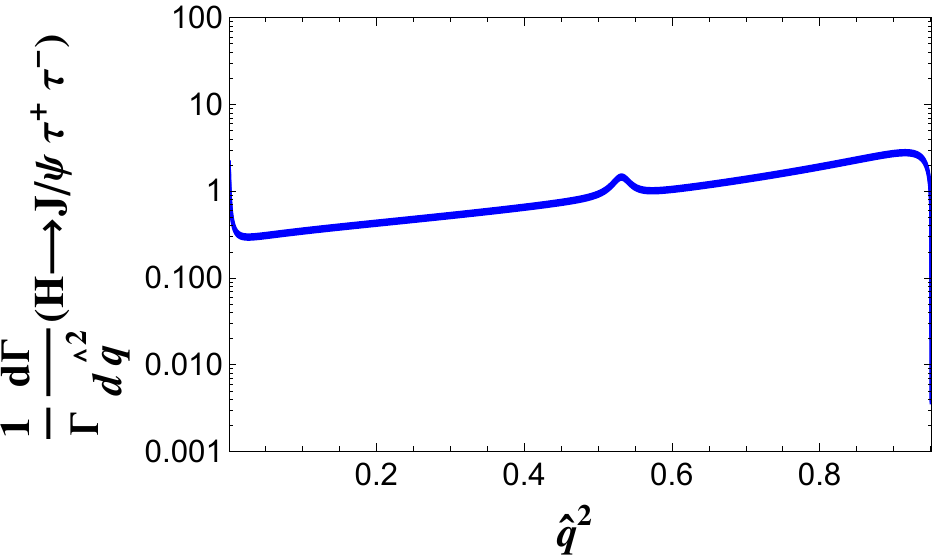}
\includegraphics[width = .3\textwidth]{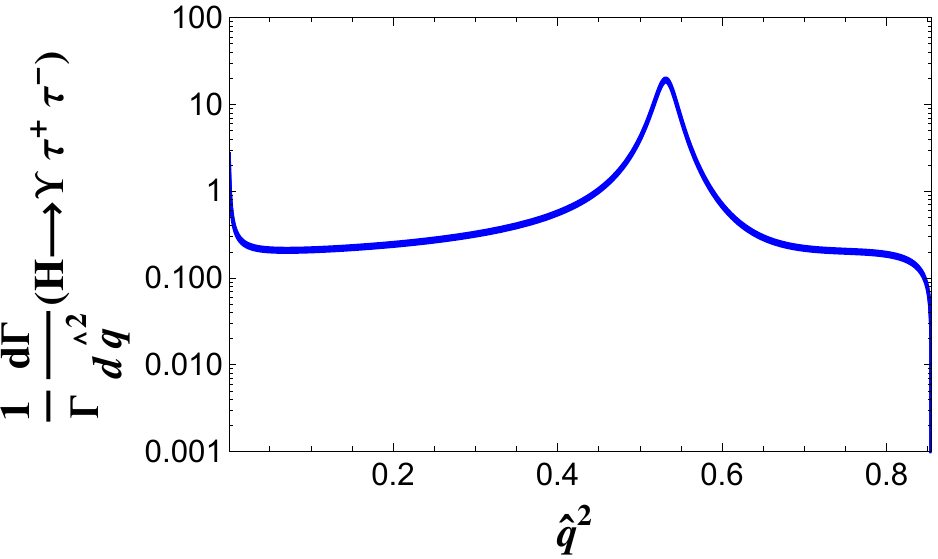}
\caption{The decay distributions $(1/\Gamma) d {\Gamma}(h \to V \ell^+ \ell^-)/d \hat q^2$, with $\hat q^2= q^2/m_h^2$ and $q^2$ the dilepton mass squared.}
\label{fig:distributions}
\end{figure}

The larger rates in Eq. \eqref{br} are predicted  for  modes with $\tau$ pairs,  $h \to \phi \tau^+  \tau^-$ and $h \to J/\psi \tau^+  \tau^-$. 
Due to the smaller coupling, the processes with muons have rates suppressed by a factor 30 and 20, respectively;
however, the higher experimental identification efficiency should cancel this suppression.

In the case of $\Upsilon$, the branching ratios with $\tau^+  \tau^-$ and $\mu^+  \mu^-$ are similar. This is an effect of
the dominance of the diagram with two intermediate $Z$, Fig.~\ref{fig:diagrams}~(c), and with coincident contributions. 
For the $h \to \Upsilon \mu^+ \mu^-$ the next most relevant diagram is the one with the Higgs coupled to quarks,  
Fig.~\ref{fig:diagrams}~(a), and it gives contribution very similar to the one coming from the diagram with the Higgs 
coupled to leptons,  Fig.~\ref{fig:diagrams}~(b), which has the same role for the $h \to \Upsilon \tau^+ \tau^-$ mode.

A comparison with the branching ratios of two body modes is in order.  In \cite{Koenig:2015pha} the authors found
${\cal B}(h\to \phi \gamma) = (2.31 \pm 0.11)\times 10^{-6}$ and
${\cal B}(h\to J/\psi \gamma) = (2.95 \pm 0.17)\times 10^{-6}$,  while   ${\cal B}(h\to \Upsilon \gamma)$ 
is   ${\cal O}(10^{-9})$. For the  $h\to V Z$  modes,   ${\cal B}(h \to \phi Z) \simeq  {\cal B}(h\to J/\psi Z) = 2.2\times 10^{-6}$ 
are  expected in SM  \cite{Isidori:2013cla}.

In Fig.~\ref{fig:distributions} the decay distributions are plotted in the normalized dilepton mass squared $\hat q^2=q^2/m_h^2$.
The studied modes, with the exception of $h \to \phi  \tau^+ \tau^-$ and $h \to J/\psi  \tau^+ \tau^-$, are dominated by 
the virtual photon and  $Z$ contributions in  Fig.~\ref{fig:diagrams}~(c). In the $h \to \phi  \tau^+ \tau^-$ and 
$h \to J/\psi  \tau^+ \tau^-$ the $\hat q^2$-distributions show a small $Z$ peak and increase with $\hat q^2$: 
an effect of the diagrams with the Higgs  coupled to the leptons.\\
For all the modes the forward-backward lepton asymmetry is very small in the whole range of $\hat q^2$. 

\begin{figure}
\centering
\includegraphics[width = .3\textwidth]{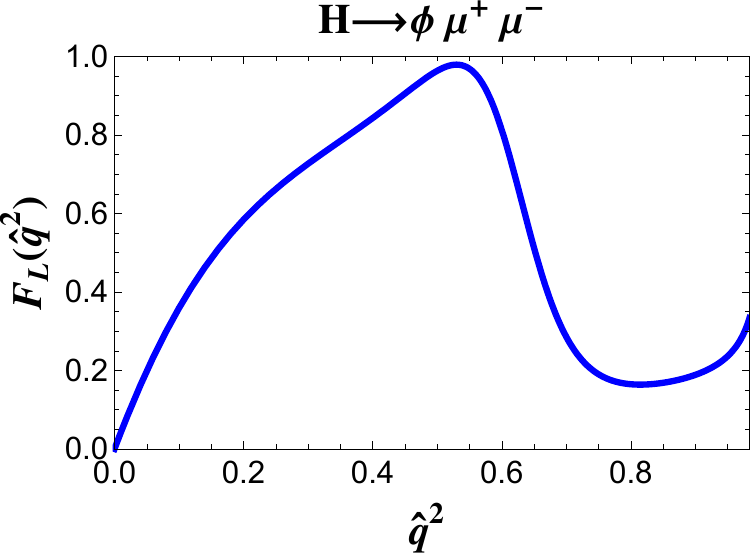}
\includegraphics[width = .3\textwidth]{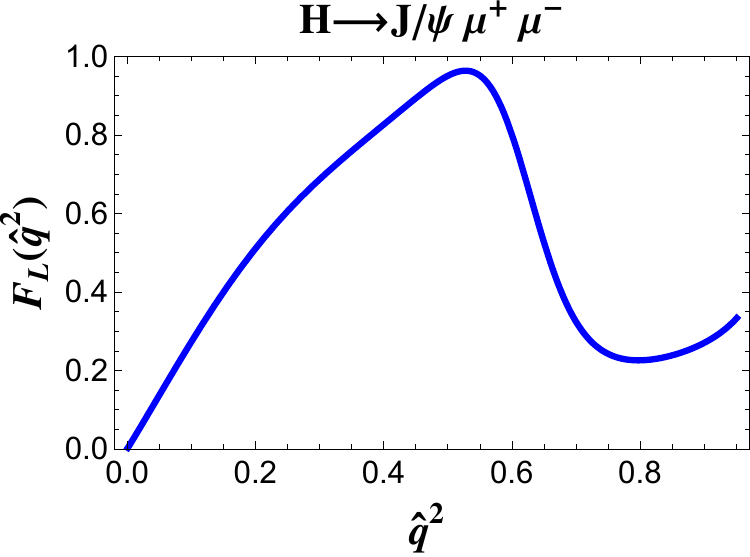}
\includegraphics[width = .3\textwidth]{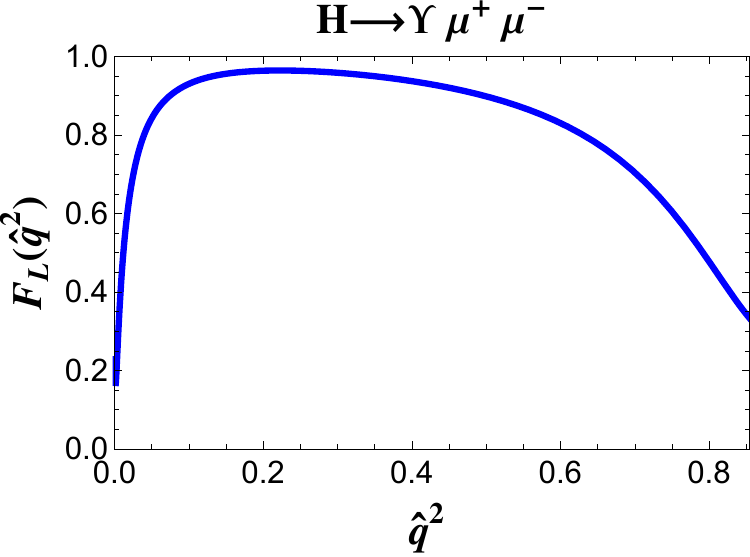}
\\ 
 \includegraphics[width = .3\textwidth]{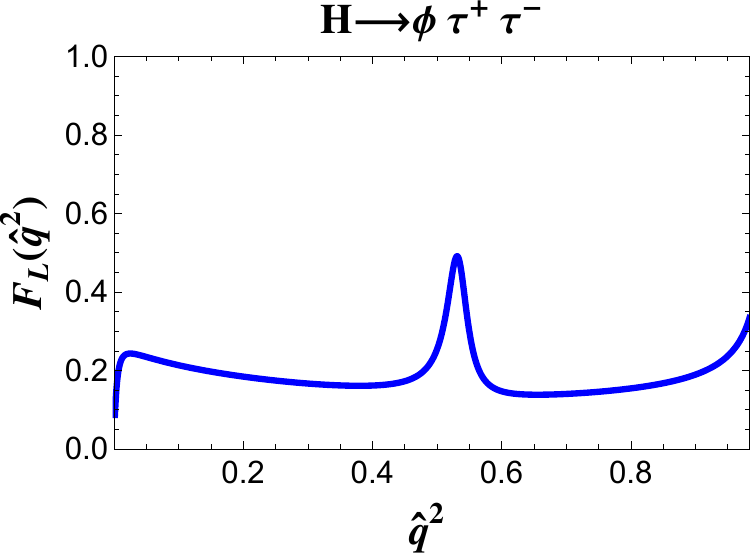}
\includegraphics[width = .3\textwidth]{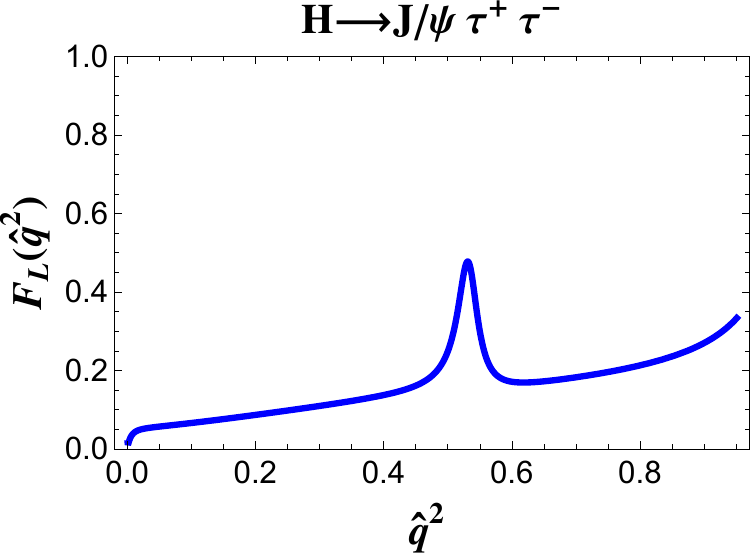}
\includegraphics[width = .3\textwidth]{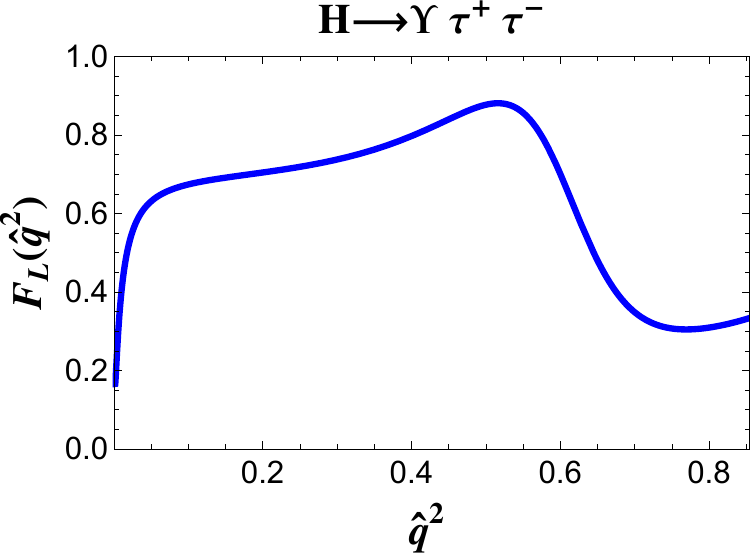}
\caption{Fraction $F_L(\hat q^2)$ of longitudinally polarized meson.}
\label{fig:long}
\end{figure}

In Fig.~\ref{fig:long}, the 
$F_L(\hat q^2)=\left(d \Gamma_L (h \to V \ell^+ \ell^-)/d \hat q^2\right)/\left(d \Gamma(h \to V \ell^+ \ell^-)/d \hat q^2\right)$
distributions of the fraction of longitudinally polarised vector meson are depicted.
$F_L\simeq 1$  at the $Z$ mass for the modes with muons in the final state. For the $\phi \tau^+\tau^-$ and 
 $J/\psi \tau^+\tau^-$ one can see a narrow peaks in $F_L$ for  $\hat q^2 = m_Z^2/m_h^2$, all the other cases 
 present a smooth $\hat q^2$ dependence. 
It is quite simple to modify the diagrams in Fig.\ref{fig:diagrams} to be able to study  the $h \to V \nu \bar \nu$  decay widths.
Our predictions are the following
\bea
\centering
\hspace{3cm}
{\cal B}(h \to \phi \nu \bar \nu) & = & (1.50  \pm 0.075 ) \times 10^{-7}\,, \nn\\
{\cal B}(h \to J/\psi \nu \bar \nu) &  = & (1.54 \pm 0.085 ) \times 10^{-7} \,, \\
{\cal B}(h \to \Upsilon \nu \bar \nu) &  = &(1.52 \pm 0.08 ) \times 10^{-6} \, ,\nn 
\label{htonu} 
\eea
with a factor $3$ included to account for the neutrino species. The measurements of these decay modes are particularly challenging.

\begin{figure}
\centering
\includegraphics[width = .3\textwidth]{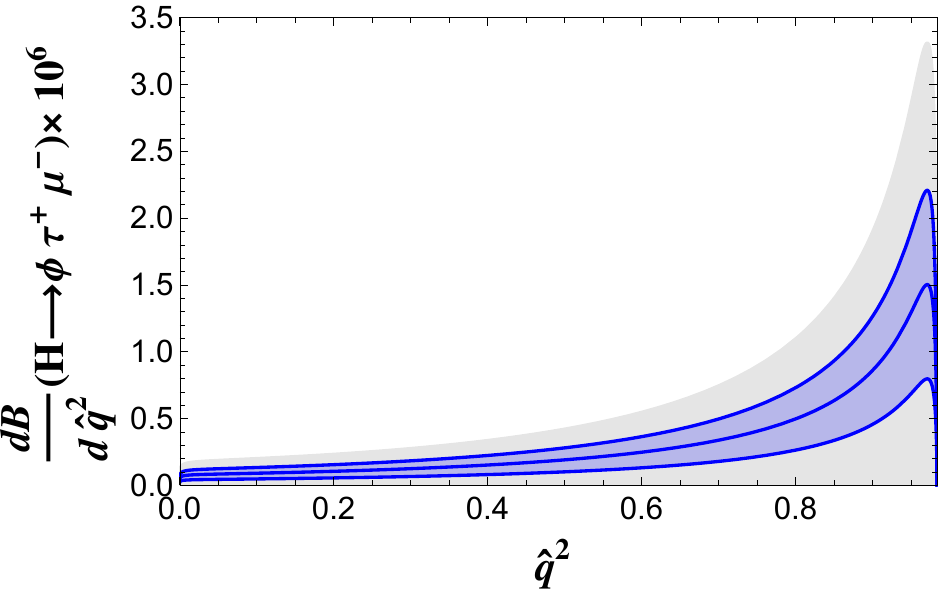}
\includegraphics[width = .3\textwidth]{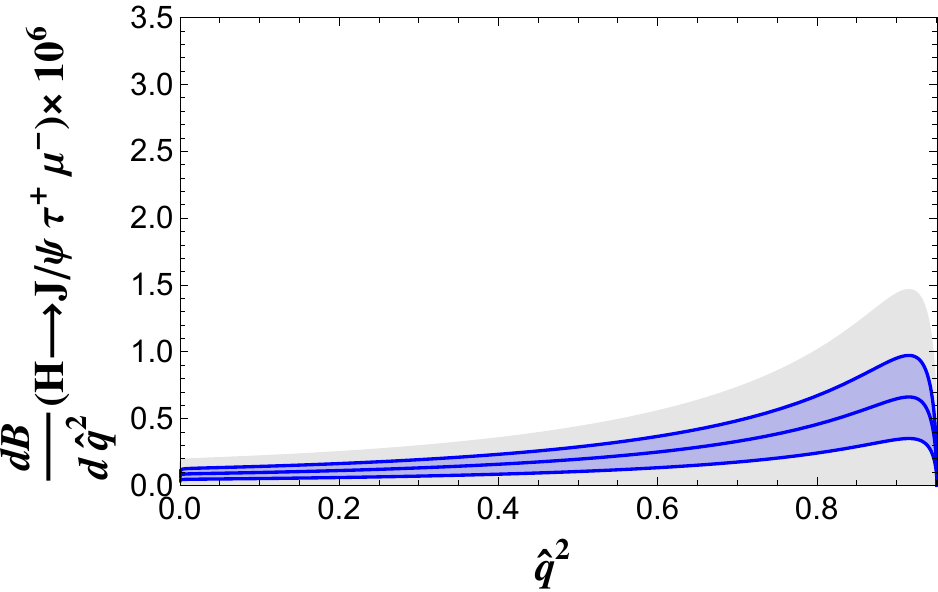}
\includegraphics[width = .3\textwidth]{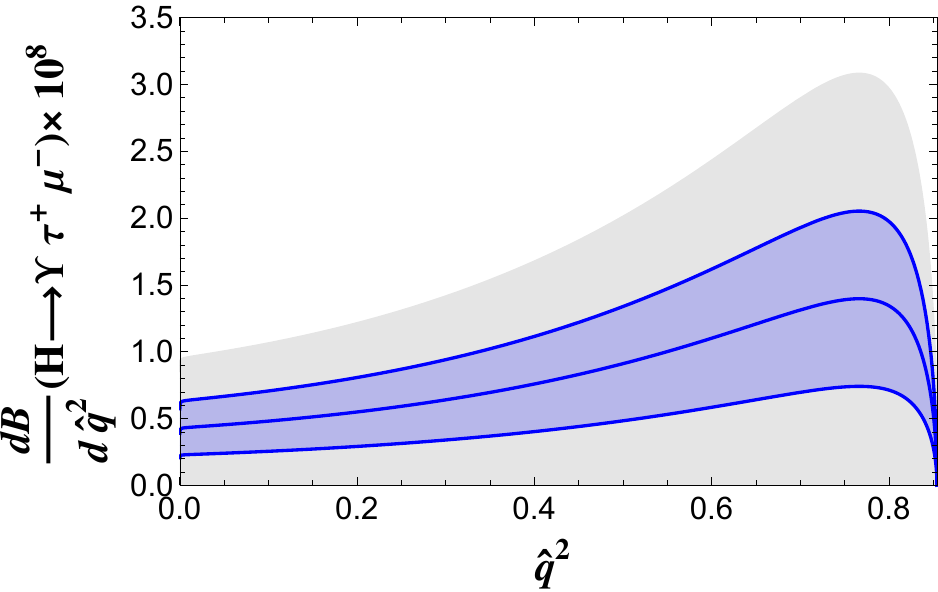}
\caption{Distributions $d {\cal B}(h \to V \tau^+ \mu^-)/d \hat q^2$ obtained in correspondence to the CMS result for 
$ {\cal B}(h \to  \tau^+ \mu^-)$ \cite{Khachatryan:2015kon}. The light shaded area corresponds to the ATLAS bound 
in \cite{Aad:2015gha}.}\label{fig:hVtaumu}
\end{figure}

Finally, it is interesting to look at the implications on the processes we have studied of possible lepton flavour violating 
transition $h \to \tau  \mu$. The $h \to \tau \mu$ process has been studied at LHC:  for such a mode the CMS Collaboration 
has published  ${\cal B}(h \to \tau \mu)=\left(0.84^{+0.39}_{-0.37}\right)\times 10^{-2}$  together with the upper bound 
${\cal B}(h \to \tau \mu)<1.51 \times 10^{-2}$ at $95\%$ CL \cite{Khachatryan:2015kon}, while the ATLAS
Collaboration quotes the bound  ${\cal B}(h \to \tau \mu)<1.85\times 10^{-2}$ at $95\%$ CL \cite{Aad:2015gha}.
By using the CMS results,  the effective coupling, $\kappa_{h \tau \mu}$, can be extracted 
$\kappa_{h \tau \mu}=(2.6 \pm 0.6)\times 10^{-3}$, considering the uncertainties on ${\cal B}(h \to \tau^+ \mu^-)$ and 
${\cal B}(h \to \gamma \gamma)$. The ATLAS upper bound, instead,  implies $\kappa_{h \tau \mu}< 3.9 \times 10^{-3}$. 
For these values, the $h \to V \tau^+ \mu^-$ branching fractions 
and their upper bounds can be computed from the diagrams in Fig.~\ref{fig:diagrams}~(b):
\bea
\hspace{2cm}
{\cal B}(h \to \phi \tau^+ \mu^-) & = &  (3.2 \pm 1.5) \times 10^{-7} \,\,\, (< 6.9 \times 10^{-7} )\,,\nn \\
{\cal B}(h \to J/\psi \tau^+ \mu^-) & = & (2.4 \pm 1.1) \times 10^{-7} \,\,\, (< 5.2 \times 10^{-7} )\,,  \\
{\cal B}(h \to \Upsilon \tau^+ \mu^-) & = & (7.2 \pm 3.4) \times 10^{-9} \,\,\, (< 1.6 \times 10^{-8} ) \, .\nn 
\label{LFV}
\eea
As one can see by looking at  Fig.~\ref{fig:hVtaumu}, all the decay distributions have an enhancement at  large  $q^2$.
\section{Conclusions}
For the exclusive decay modes $h \to V \ell^+ \ell^-$  the obtained branching ratios are in the range  $10^{-8}\div 10^{-6}$ in SM, 
of the same order of magnitude of    $h\to \phi\gamma$, $h\to \Upsilon \gamma$,  $h\to (\phi,J/\psi) \, Z$. 
The largest rate is predicted for $h\to \phi \tau^+\tau^-$.
The branching ratios of the neutrino modes have  been calculated. 
We have also studied the lepton flavour-changing process $h \to V \tau \mu$ by using the CMS and ATLAS experimental results on the 
$h \to \tau \mu$ process.

\section*{Acknowledgements}
\noindent I would like to thank Pietro Colangelo and Fulvia De Fazio for the pleasant collaboration.

\bibliography{Proceedings.bib}

\begin{thebibliography}{31}

\bibitem{Aad:2012tfa}
G.~Aad et~al. (ATLAS), Phys. Lett. \textbf{B716}, 1 (2012), \texttt{1207.7214}

\bibitem{Chatrchyan:2012xdj}
S.~Chatrchyan et~al. (CMS), Phys. Lett. \textbf{B716}, 30 (2012),
  \texttt{1207.7235}

\bibitem{Agashe:2014kda}
K.A. Olive et~al. (Particle Data Group), Chin. Phys. \textbf{C38}, 090001
  (2014)

\bibitem{atlascms}
ATLAS, CMS, ATLAS-CONF-2015-044, CMS-PAS-HIG-15-002 (2015)

\bibitem{Contino:2013kra}
R.~Contino, M.~Ghezzi, C.~Grojean, M.~Muhlleitner, M.~Spira, JHEP \textbf{07},
  035 (2013), \texttt{1303.3876}

\bibitem{Brivio:2013pma}
I.~Brivio, T.~Corbett, O.J.P. \'{E}boli, M.B. Gavela, J.~Gonzalez-Fraile, M.C.
  Gonzalez-Garcia, L.~Merlo, S.~Rigolin, JHEP \textbf{03}, 024 (2014),
  \texttt{1311.1823}

\bibitem{Gonzalez-Alonso:2014eva}
M.~Gonzalez-Alonso, A.~Greljo, G.~Isidori, D.~Marzocca, Eur. Phys. J.
  \textbf{C75}, 128 (2015), \texttt{1412.6038}

\bibitem{Gupta:2014rxa}
R.S. Gupta, A.~Pomarol, F.~Riva, Phys. Rev. \textbf{D91}, 035001 (2015),
  \texttt{1405.0181}

\bibitem{Chien:2015xha}
Y.T. Chien, V.~Cirigliano, W.~Dekens, J.~de~Vries, E.~Mereghetti (2015),
  \texttt{1510.00725}

\bibitem{Abbasabadi:1996ze}
A.~Abbasabadi, D.~Bowser-Chao, D.A. Dicus, W.W. Repko, Phys. Rev. \textbf{D55},
  5647 (1997), \texttt{hep-ph/9611209}

\bibitem{Chen:2012ju}
L.B. Chen, C.F. Qiao, R.L. Zhu, Phys. Lett. \textbf{B726}, 306 (2013),
  \texttt{1211.6058}

\bibitem{Sun:2013rqa}
Y.~Sun, H.R. Chang, D.N. Gao, JHEP \textbf{05}, 061 (2013), \texttt{1303.2230}

\bibitem{Dicus:2013ycd}
D.A. Dicus, W.W. Repko, Phys. Rev. \textbf{D87}, 077301 (2013),
  \texttt{1302.2159}

\bibitem{Passarino:2013nka}
G.~Passarino, Phys. Lett. \textbf{B727}, 424 (2013), \texttt{1308.0422}

\bibitem{Bodwin:2013gca}
G.T. Bodwin, F.~Petriello, S.~Stoynev, M.~Velasco, Phys. Rev. \textbf{D88},
  053003 (2013), \texttt{1306.5770}

\bibitem{Kagan:2014ila}
A.L. Kagan, G.~Perez, F.~Petriello, Y.~Soreq, S.~Stoynev, J.~Zupan, Phys. Rev.
  Lett. \textbf{114}, 101802 (2015), \texttt{1406.1722}

\bibitem{Isidori:2013cla}
G.~Isidori, A.V. Manohar, M.~Trott, Phys. Lett. \textbf{B728}, 131 (2014),
  \texttt{1305.0663}

\bibitem{Koenig:2015pha}
M.~Koenig, M.~Neubert, JHEP \textbf{08}, 012 (2015), \texttt{1505.03870}

\bibitem{Bhattacharya:2014rra}
B.~Bhattacharya, A.~Datta, D.~London, Phys. Lett. \textbf{B736}, 421 (2014),
  \texttt{1407.0695}

\bibitem{Gao:2014xlv}
D.N. Gao, Phys. Lett. \textbf{B737}, 366 (2014), \texttt{1406.7102}

\bibitem{Colangelo:2016jpi}
P.~Colangelo, F.~De~Fazio, P.~Santorelli, Phys. Lett. \textbf{B760}, 335
  (2016), \texttt{1602.01372}

\bibitem{Lepage:1979zb}
G.P. Lepage, S.J. Brodsky, Phys. Lett. \textbf{B87}, 359 (1979)

\bibitem{Lepage:1980fj}
G.P. Lepage, S.J. Brodsky, Phys. Rev. \textbf{D22}, 2157 (1980)

\bibitem{Efremov:1979qk}
A.V. Efremov, A.V. Radyushkin, Phys. Lett. \textbf{B94}, 245 (1980)

\bibitem{Chernyak:1983ej}
V.L. Chernyak, A.R. Zhitnitsky, Phys. Rept. \textbf{112}, 173 (1984)

\bibitem{Grossmann:2015lea}
Y.~Grossman, M.~Koenig, M.~Neubert, JHEP \textbf{04}, 101 (2015),
  \texttt{1501.06569}

\bibitem{Caswell:1985ui}
W.E. Caswell, G.P. Lepage, Phys. Lett. \textbf{B167}, 437 (1986)

\bibitem{Bodwin:1994jh}
G.T. Bodwin, E.~Braaten, G.P. Lepage, Phys. Rev. \textbf{D51}, 1125 (1995),
  [Erratum: Phys. Rev.D55,5853(1997)], \texttt{hep-ph/9407339}

\bibitem{Heinemeyer:2013tqa}
J.R. Andersen et~al. (LHC Higgs Cross Section Working Group) (2013),
  \texttt{1307.1347}

\bibitem{Khachatryan:2015kon}
V.~Khachatryan et~al. (CMS), Phys. Lett. \textbf{B749}, 337 (2015),
  \texttt{1502.07400}

\bibitem{Aad:2015gha}
G.~Aad et~al. (ATLAS), JHEP \textbf{11}, 211 (2015), \texttt{1508.03372}

\end{thebibliography}

\end{document}